# Controlled Growth of large area bilayer MoS$_2$ films on SiO$_2$ substrates by chemical vapour deposition technique

Umakanta Patra, Faiha Mujeeb, Abhiram K, Jai Israni, Subhabrata Dhar*

Department of physics, Indian institute of technology, Bombay, Mumbai, India,400076

Corresponding author email: dhar@phy.iitb.ac.in

Bilayer (2L) transition metal dichalcogenides (TMD) have the ability to host interlayer excitons, where electron and hole parts are spatially separated that leads to much longer lifetime as compared to direct excitons. This property can be utilized for the development of exciton-based logic devices, which are supposed to be superior in terms of energy efficiency and optical communication compatibility as compared to their electronic counterparts. However, obtaining uniformly thick bilayer epitaxial films with large area coverage is challenging. Here, we have engineered the flow pattern of the precursors over the substrate surface to obtain large area (mm$^2$) covered strictly bilayer MoS$_2$ films on SiO$_2$ by chemical vapour deposition (CVD) technique without any plasma treatment of the substrate prior to the growth. Bilayer nature of these films is confirmed by Raman, low-frequency Raman, atomic force microscopy (AFM) and photoluminescence (PL) studies. The uniformity of the film has been checked by Raman peak separation and PL intensity map. High resolution transmission electron microscopy (HRTEM) reveals that crystalline and twisted bilayer islands coexist within the layer. Back gated field-effect transistor (FET) structures fabricated on the bilayers show on/off ratio of $10^6$ and subthreshold swings (SS) of 2.5 V/Decade.

**Keywords:** Bilayer MoS$_2$, Large area, CVD, Field effect transistors

## Introduction

Monolayer (1L) transition metal dichalcogenides (1L-TMDCs) exhibit certain unique properties such as the large exciton binding energy [1], existence of many body excitonic states, e.g. trions, biexcitons [2–4] etc. and most notably the valley polarization (VP) that results from the broken inversion symmetry and strong spin-orbit coupling in the material [5,6]. This enables an exciton in K (or K')-valleys to sustain its valley character throughout the time of its existence. Moreover, the spin state of the exciton is tagged with the valley index. The property can be exploited to develop valley-based electronics or 'valleytronics', which can be used for the development of multifunctional quantum devices, and quantum computation [7,8]. Another important aspect of this class of 2D materials is layer number dependent electronic properties. For example, monolayer TMDCs have direct band gap, while multilayers show indirect nature [9,10]. One Interesting feature of bilayer TMDCs is its ability to host interlayer excitons [11]. These excitons, where electron and hole parts are spatially separated, have much longer lifetime as compared to direct excitons [12]. This aspect can be used for the development of exciton-based logic devices, which have superior energy efficiency and optical communication compatibility compared to their electronic counterparts [13,14]. Moreover, due to smaller thickness, high stiffness, breaking strength, lower dielectric constant, higher electron effective mass and larger band gap than silicon, TMDC films are superior choice for the suppression of the short channel effects to achieve larger density of devices in a chip [15]. In this respect, TMDC films with thickness of a few layers are preferred over monolayers as the carrier mobility in the former is expected to be more than in the latter due to lower surface to volume ratio [16]. It should be further mentioned that TMDCs have shown great potential for neuromorphic computing applications [17]. There are several reports on artificial synaptic devices based on these materials [18]. Recently, M.M Islam *et al.* has demonstrated self-denoising capability of the chemical vapor deposition (CVD) grown bilayer MoS$_2$ based artificial optoelectronic synapse [19]. However, in most of the cases, these devices are demonstrated on isolated bilayer islands. This restricts the fabrication of multiple devices on a single platform necessary for the realization of the artificial neural network [20]. MoS$_2$ is the most studied material within the class of 2D TMDCs [21]. Various growth methods, such as laser thinning [22], liquid exfoliation [23] and hydrothermal synthesis [24], are adopted to grow bilayer MoS$_2$. However, not much control over the area coverage and the layer numbers [25] could be achieved by these methods. CVD technique has often been used to grow large area covered continuous 1L MoS$_2$ films [26,27]. Lei Liu *et al.* has reported the epitaxial growth of continuous 2L-MoS$_2$ films covering large area (few cm$^2$) on c-plane sapphire substrates using the same technique [28]. However, growing large area continuous films of bilayer (2L) MoS$_2$ on SiO$_2$/Si substrates is still a challenge. Note that large area films are important for any practical application [29]. Further for better integration with existing Si-technology, it is necessary that these films are grown on SiO$_2$/Si substrates [30]. Recently, Xiumei Zhang *et al.* has reported the growth of 2L MoS$_2$ films on SiO$_2$/Si substrates by controlling the growth temperature and the gas flow direction in chemical vapour deposition (CVD) [31]. However, bilayers grown by this method are found to be isolated domains of size not more than a few hundreds of micrometer. Tingting Xu *et al.* reported the growth of 2L MoS$_2$ triangles of size 50-100 $\mu$m on SiO$_2$/Si by a two-step CVD process, where the MoO$_3$ film is first deposited on the substrate and then sulphurisation of the film is carried out [32]. Sungjoo Lee *et al.* used a low-pressure chemical vapour deposition technique (LPCVD) to grow continuous MoS$_2$ films with different layer numbers on SiO$_2$ substrate treated with O$_2$ plasma for different times prior to the growth. However, the morphology of SiO$_2$ film has been found to deteriorate after the plasma treatment. Note that this deterioration is unwanted keeping in mind the need for low leakage current in any electronic device developed on this platform [33]. S.Xiao *et al.* has reported CVD growth of continuous monolayer, bilayer and multilayer MoS$_2$ films on SiO$_2$/Si substrate by varying the growth temperature. However, a details investigation on the bilayer nature of the MoS$_2$ films is lacking in their study. [34]

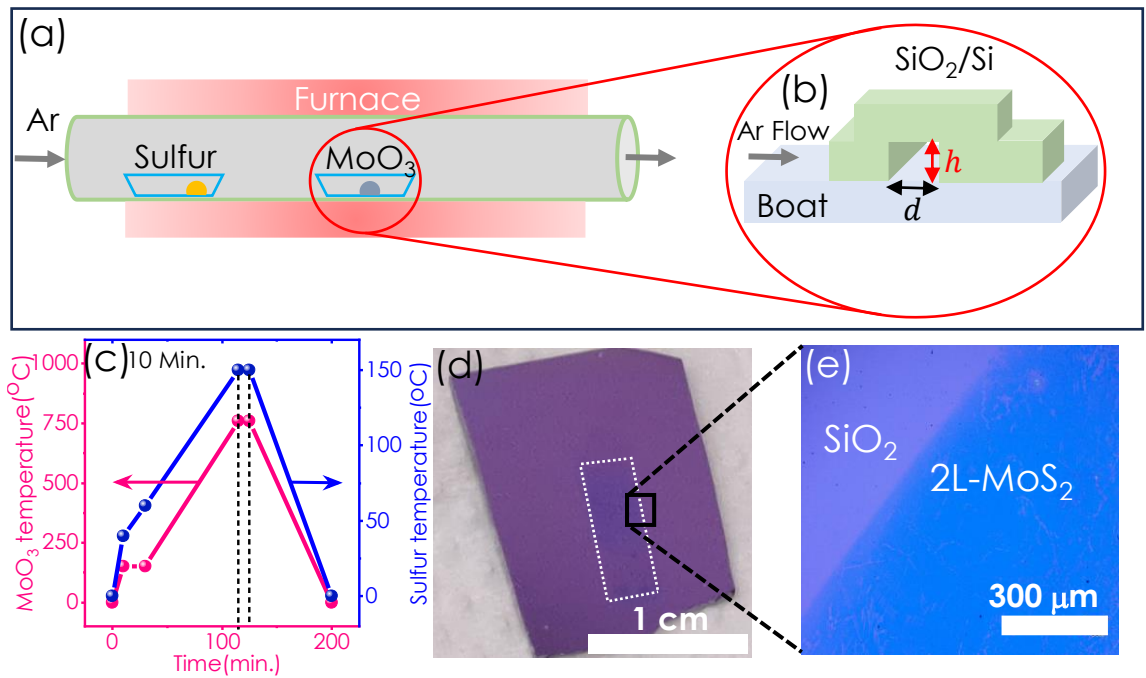

**Figure. 1.** (a) Chemical Vapour Deposition set up. (b) Arrangement of the substrates on top of the MoO$_3$ containing boat. (c) Temperature rise and cooling curves at the locations of MoO$_3$ and S powder containing boats. (d) Photograph of the substrate(roof piece) after the growth. (e) Zoomed-in view of the portion of the substrate, which was directly exposed to MoO$_3$ powder. Continious growth of MoS$_2$ layer (blue area) could be seen.

Here, we have engineered the flow pattern of the sulphur and MoO$_3$ vapours over the substrate surface by introducing blockades in a CVD setup to achieve the growth of large area continuous 2L-MoS$_2$ films on $SiO_2$/Si substrates. We have optimized the growth temperature, amount of the precursors as well as the width and height of the vapor flow channel to control the reaction rate of the precursors on the substrate surface. It has been found that at the optimum condition, $MoS_2$ films with the thickness of 2 monolayers (2L) can be deposited uniformly on $SiO_2$/Si substrates covering an area as large as several tens of $mm^2$. We have also carried out the growth of $MoS_2$ on $SiO_2$ after treating the $SiO_2$/Si substrates in $O_2$-plasma for different amounts of time. It has been found that the plasma exposure does not favour the growth of 2L-$MoS_2$. Instead, 1L-$MoS_2$ films are deposited under the same growth condition even for a few seconds of plasma treatment. The findings thus provide a novel way to grow a large area covered strictly bilayer $MoS_2$ films on $SiO_2$/Si substrate, which can be useful for developing 2D materials-based electronic/excitonic devices. Back gated field-effect transistor (FET) structures are fabricated on these bilayers, which show on/off ratio as high as $10^6$, mobility of 9.5 $cm^2$/V-s and subthreshold swings of 2.5 V/decade.

**Experimental methods**

**Growth and characterization of 2L-$MoS_2$**

Samples were grown in a CVD chamber as depicted schematically in Fig. 1(a). Two ceramic boats, one containing 2 mg of high purity $MoO_3$ powder (99.995 %, Aldrich) and another with 350 mg of sulphur (S) powder (99.95%, Aldrich), were kept respectively in the higher and lower temperature zones of the quartz reactor tube. Separation between two boats was maintained at 16.5 cm. The $SiO_2$/Si substrate was cleaned successively in tri chloroethylene (TCE), acetone and methanol for five minutes each in an ultrasonic bath before placing it on top of the $MoO_3$ containing boat as shown in the Fig. 1(b). The quartz reactor was purged with argon (Ar) gas for 10 min at a flow rate ($\varphi_{Ar}$) of 300 sccm at room temperature. The furnace temperature is then raised to 150℃ and Ar-purging was continued for another 20 minutes at $\varphi_{Ar}$ = 200 sccm. $\varphi_{Ar}$ was then reduced to 8 sccm while the furnace temperature was raised to 700℃ at a rate of 10℃/min. This temperature was maintained for 10 minutes, which allowed the sulphur boat temperature to reach 150℃. The furnace was then switched off to cool down the substrate naturally under the flow of argon. Temperature evolution profiles at the $MoO_3$ and sulphur boat positions are shown in the Fig. 1(c). Photographs of the substrate(roof piece) after the growth is shown in Fig. 1(d) and (e), which show the deposition of a continuous $MoS_2$ film (blue area). Note that the separation between the supporting pieces $d$ and the height of the substrate $h$ [see Fig. 1(b)] play very crucial roles in governing the size of the bilayer islands [see supplementary information S.1]. It should be noted that the flow pattern of the precursors plays a vital role in the growth process. M A Gokul et al. have manipulated the growth pattern by placing blockades of different size and shape on the substrates during the CVD growth of their $MoS_2$ layers [35]. They have found the growth of monolayer $MoS_2$ only in the regions, where the flow trabulance is supressed. Here the we believe that for a particluar the height and width of the channel in our set up the velocity distribution of the precursor particles becomes uniform. This laminar flow results in the growth of uniform deposition of $MoS_2$.

$MoS_2$ films were also grown on $SiO_2$ after treating the surface in $O_2$-plasma for 10, 30, 60 and 180 sec of exposer times. The plasma treatment was carried out at chamber pressure of 0.5 mbar, $O_2$ flow rate of 5 sccm and plasma power of 50 W. It has been found that $O_2$ plasma treated substrates facilitate the growth of monolayer instead of bilayer $MoS_2$ films. This is discussed in supplementary section S.2.

Optical and atomic force microscopy (AFM) of the sample were carried out at room temperature. Photoluminescence (PL) and Raman spectroscopy were performed using a micro-PL/Raman setup from Renishaw, Invia Reflex, UK. PL and Raman maps were recorded with a 50× magnification and 1$\mu$m of scanning step. Low frequency Raman and PL spectroscopy are carried out using a 532 nm diode laser excitation source with 500$\mu$W power and a 500 cm focal-length monochromator equipped with 2400 gr/mm grating and CCD detector. High resolution transmission electron microscopy (HRTEM) investigation is carried out in a 300 kV system from Thermo Scientific (Themis 300 G3) after transferring the film onto a Cu-grid using polystyrene based wet transfer technique [36].

**Device fabrication and characterization**

Bilayer $MoS_2$ films were transferred on a 300 nm thick $SiO_2$ coated p-type Si-substrates by polystyrene based wet transfer technique [36]. Backgated FET devices were fabricated using standard optical lithography and lift-off technique. Deposition of Ag(30nm) and Au(100nm) was carried out by sputtering techniques. Keithley 6487 pico-ammeter-voltage source and a Keysight B2901B source measuring unit were used for the electrical characterizations.

**Result and Discussion**

Room temperature Raman spectrum of the film is shown in Fig. 2(a). Raman spectrum for a 1L-MoS$_2$, which has been grown in the same CVD setup, is shown for comparison. The characteristic $E_{2g}^1$ (In-plane vibration) and $A_{1g}$ (out-of plane vibration) peaks of MoS$_2$ could be seen for both the cases. The seperation between $E_{2g}^1$ and $A_{1g}$ peaks, which carries the information of the layer thickness, is found to be 21 and 19 cm$^{-1}$, respectively, for this film and the monolayer 1L-MoS$_2$ film. The increased raman seperation indicates that the film under study must be bilayer MoS$_2$ [16,37,38]. The map showing the separation between $E_{2g}^1$ and $A_{1g}$ peaks is shown in the inset of Fig. 2(a). Evidently, the gap between $E_{2g}^1$ and $A_{1g}$ does not show much variation over the entire scanned area of 900$\mu m^2$. This suggests that the bilayer nature of the deposition is maintained over a large area.

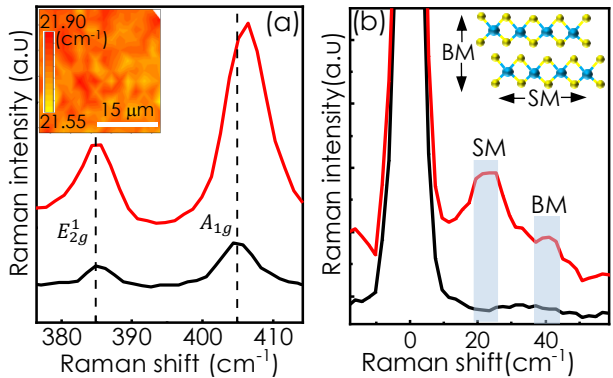

**Figure. 2.** (a) Room temperature Raman spectra comparing monolayer (black line) and bilayer (red line) MoS$_2$. Inset shows the $E_{2g}^1$-$A_{1g}$ seperation map. (b) Low frequency Raman spectra recorded for monolayer (black line) and bilayer (red line) MoS$_2$. Inset schematically represents the shear (SM) and breathing (BM) modes.

Figure 2(b) compares low frequency raman spectra for the film and the 1L- MoS$_2$. Presence of two low frequency raman features at 23 and 40 cm$^{-1}$ is quite evident in the studied MoS$_2$ film. While these features are missing in the 1L-MoS$_2$. The two peaks can be attributed to the in-plane shear mode (SM) [at 23cm$^{-1}$] and out-of-plane breathing mode (BM) [at 40 cm$^{-1}$] observed in bilayer MoS$_2$ [37]. Note that the two modes should not be present in 1L-MoS$_2$, which is consistant with the observation. The finding thus confirms the bilayer nature of of our MoS$_2$ film.

Figure 3(a) compares the room temperature photoluminoscence(PL) spectra recorded on 1L- (black symbols) and our 2L-MoS$_2$ (orange symbol) films. In both the cases, the spectra are featured by two peaks, which can be assigned to A and B excitons in the K-valleys. Evidently, the A-excitonic peak intensity is suppressed in our 2L-MoS$_2$ film as compared to 1L-MoS$_2$. While, the intensity of the B excitonic feature is not so different in the two cases. Such a dramatic reduction of A-excitonic intensity in 2L- as compared to 1L-MoS$_2$ has been reported earlier and can be attributed to the formation of spatially indirect excitons in the former [39]. Fig. 3(b) presents the PL intensity map for the MoS$_2$ film over a 20 × 20 um$^2$ area. The bare substrate region in the image is marked by white dotted line. Intenisty is evidently uniform over most parts of the selected region, which further confirm the uniformity of the 2L-MoS$_2$ film.

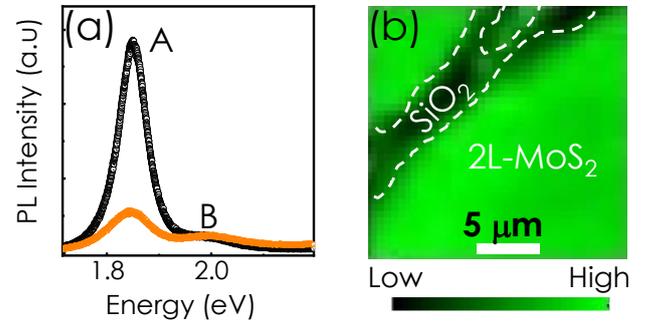

**Figure. 3.** (a) Room temperature PL spectra recorded on the monolayer (Black) and bilayer (Orange) MoS$_2$. (b) Room temperature PL intensity map.

The layer thickness is further explored by atomic force microscopy (AFM). Fig. 4(a) is showing the AFM image taken from the edge of the continious MoS$_2$ film. Fig. 4(b) shows the magnified view of the marked region [white box] of the AFM image of Fig. 4(a). Presence of two layers with a clear side step is evident in the image. The line-scan profile along the white line marked in Fig. 4(b) is shown in the the inset of the figur. The profile clearly shows two step increase of height. The thickness is found to be 1.7 nm, which also matches well with other reported values for CVD grown bilayer MoS$_2$ thin films [38].

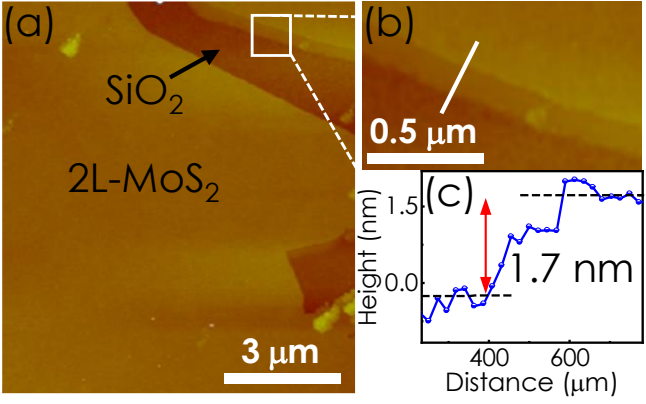

**Figure. 4.** Atomic force microscopy image of the continious bilayer MoS₂ film grown on SiO₂/Si substrate. (b) Magnified view of the edge of the film evidencing the stacking of two layers. (c) Height profile recorded along the white line maked in the figure 4(b).

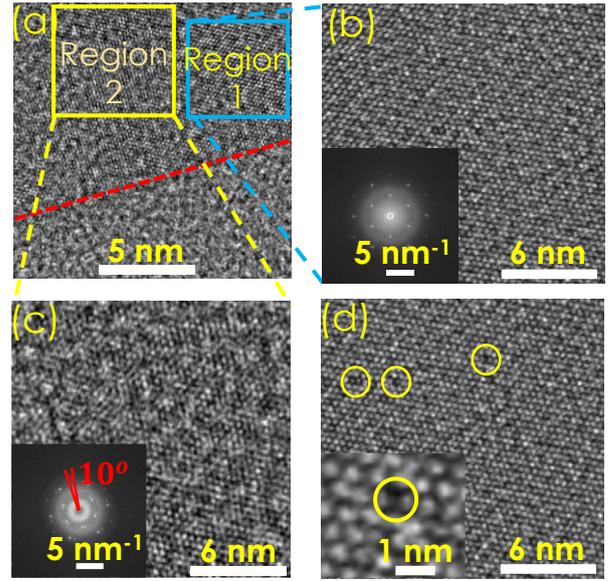

**Figure. 5.** High resolution TEM (HRTEM) images: (a) HRTEM image from a portion of the film. Region 1 and 2 are marked in the image. (b) HRTEM image of region 1 and the corresponding fast furrier transform (FFT) pattern. (c) HRTEM image of region 2 and the corresponding FFT pattern. Double spot formation in FFT indicates the presence of twisted bilayers in this part of the film. The twist angle is estimated to be 10°. (d) HRTEM image of a crystalline region, inset shows a vacancy in an expanded scale.

Figure 5(a) shows the HRTEM image of the bilayer MoS₂ film. Edge of the continuous film on the Cu-gride is marked by the red dotted line. Fig. 5(b) and (c) show the magnified views of region 1 (blue box) and region 2 (yellow box), respectively. Crystalline islands extending over several hundreds of nanometres are observed as shown in these figures. Respective insets present the fast Fourier transform (FFT) patterns of the images. Perfect hexagonal arrangement of the spots in the FFT pattern of region 1, confirms the AA stacking of the monolayers. Interestingly, the Moire pattern with a twist of 10° has been observed in the region 2 [Fig. 5(c)]. Note that similar Moire pattern has been reported earlier in 2L-MoS₂ by other group [40]. These results suggest that both AA stacked and twisted bilayer islands coexist within the sample. Presence of vacancies are also evident in the HRTEM images, which has been highlighted in Fig. 5(d).

Fig. 6(a) plots the variation of source-drain ($I_{ds}$) current as a function of source-drain ($V_{ds}$) voltage at different gate-source ($V_{gs}$) voltages. The top-view optical image of the source and drain contacts can be seen in the inset of the figure. Enhancement of $I_{ds}$ from a few pA to μA with the increase of $V_{gs}$ from -40 to 40 V confirms the n-type nature of the bilayer MoS₂ films. Fig. 6(b) shows the variation of $I_{ds}$ with $V_{gs}$ at $V_{ds} = 10$ V. From the data, the field effect mobility of the electrons can be obtained from the following equation [29].

$$\mu = \left(\frac{dI_{ds}}{dV_{gs}}\right) \times \frac{L}{c_{ox} \times W \times V_{ds}} \quad (1)$$

Where $L$ (~5 μm) is the channel length and $W$ (~5 μm) is the channel width. C$_{ox}$ is the gate capacitance, which is estimated to be 11.51 nFcm⁻² $dI_{ds}/dV_{gs} = 1.09 \times 10^{-6}$ A/V obtained from the $I_{ds}$ versus $V_{gs}$ plots shown in the inset of Fig. 6(b). From these values, the mobility $\mu$ can be estimated as ~9.5 cm²/V-s. Furthermore, the on/off ratio of the device and the subthreshold swings are estimated to be ~$10^6$ and 2.5 V/Decade, respectively. Note that in terms of these figure of merits our 2L-MoS$_2$ based FET is quite comparable with those reported on CVD grown 2L-MoS$_2$ films as shown in Table 1. Note that high values of carrier mobility have been observed only for the devices fabricated on small isolated flakes. This might be due to the lower rate of grain boundary scattering events for the carriers in these small islands. Performance of our FET devices might be further improved by introducing high K-gate dielectrics as the top gate [42,43]. Comparison of the performance of our bilayer MoS$_2$ based back gated FET has been shown in Table. 1.

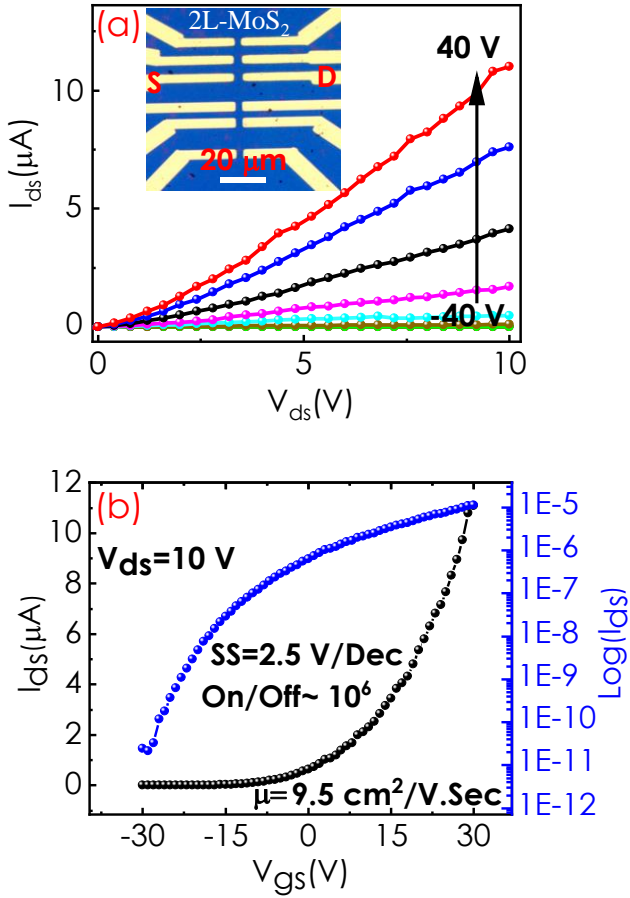

**Figure. 6.** (a) Variation of source-drain current ($I_{ds}$) as a function of source-drain voltage ($V_{ds}$) at different gate-to-source voltages ($V_{gs}$), inset shows the optical top-view image of the device (b) Variation of $I_{ds}$ as a function of $V_{gs}$ at a fixed source-drain voltage $V_{ds}$ = 10 V.

**Table. 1.** Comparison of the FET performance of CVD grown 2L-MoS$_2$

| Sl.no | Device structure | Coverage | Carrier Mobility(cm²/V.Sec) | ON/OFF ratio | SS value (V/Dec) | Ref. |
|---|---|---|---|---|---|---|
| 1 | 2L-MoS$_2$/SiO$_2$/Si | Continuous films | 8.2 | --- | --- | [33] |
| 2 | 2L-MoS$_2$/SiO$_2$/Si | Isolated flakes | 21 | $10^7$ | --- | [16] |
| 3 | 2L-MoS$_2$/SiO$_2$/Si | Isolated flakes | 3.2 | $10^5$ | --- | [44] |
| 4 | 2L-MoS$_2$/SiO$_2$/Si | Isolated flakes | 0.01 | --- | --- | [45] |
| 5 | 2L-MoS$_2$/SiO$_2$/Si | Isolated flakes | 4.5 | $10^5$ | --- | [46] |
| 6 | 2L-MoS$_2$/SiO$_2$/Si | Isolated flakes | 0.3 | $10^6$ | --- | [47] |
| 7 | 2L-MoS$_2$/HFLaO | Isolated flakes | 36 | $10^6$ | --- | [14] |
| 9 | 2L-MoS$_2$/SiO$_2$/Si | Continuous films | 9.5 | $10^6$ | 2.5 | This work |

## Conclusion

Continuous bilayer $MoS_2$ films can be grown on $SiO_2$/Si by chemical vapour deposition technique. It has been found that the flow pattern of the precursor over the substrate can be engineered to obtain large area ($mm^2$) covered strictly bilayer films without any plasma treatment of the substrate prior to the growth. HRTEM study on these films' revels the existence of both twisted and untwisted islands in the film. Back gated FET devices fabricated on these bilayer films shows excellent device characteristics.

## Acknowledgement

The author acknowledges the financial support from the Science and Engineering Research Board (SERB), under the grant number CRG/2022/001852, Government of India. The authors are also thankful to Industrial Research and consultancy Centre (IRCC), Sophisticated Analytical Instrument Facility (SAIF), and the Centre for Excellence in Nanoelectronics (CEN), IIT Bombay for providing certain characterization facilities. We also thank to Prof. Gopal K. Pradhan, KIIT deemed to be university, Bhubaneswar for providing the low-frequency Raman measurement facility.

## Conflict of interest

The author declares no conflict of interest

## Data availability statement

Data that support the findings of this study are available from the corresponding author upon reasonable request.

# Supplementary information (S.I)

# Controlled Growth of large area bilayer $MoS_2$ films on $SiO_2$ substrates by chemical vapour deposition technique

Umakanta Patra, Faiha Mujeeb, Abhiram K, Jai Israni, Subhabrata Dhar*

Department of physics, Indian institute of technology, Bombay, Mumbai, India, 400076.

S.1. Morphology of the grown layer as a function of the separation $d$ between the two supporting pieces and the height $h$ of the roof piece

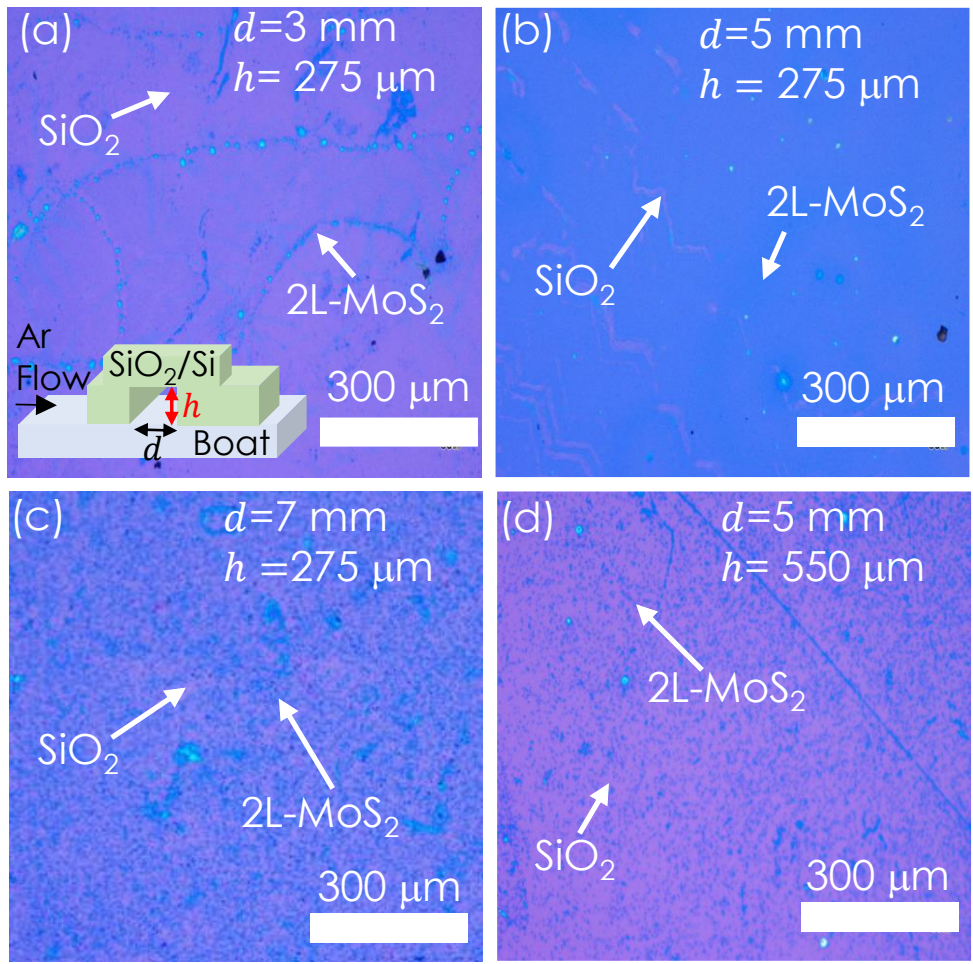

**FIG. S1.** Optical images of the samples grown with different values of $d$ (separation between the two-supporting pieces) and $h$ (height of the roof piece). Strictly 2L MoS$_2$ film covering a large area can only be grown for $d = 5$ mm and $h = 275$ $\mu$m.

S.2. Effect of O$_2$-Plasma treatment on the growth of 2L-MoS$_2$

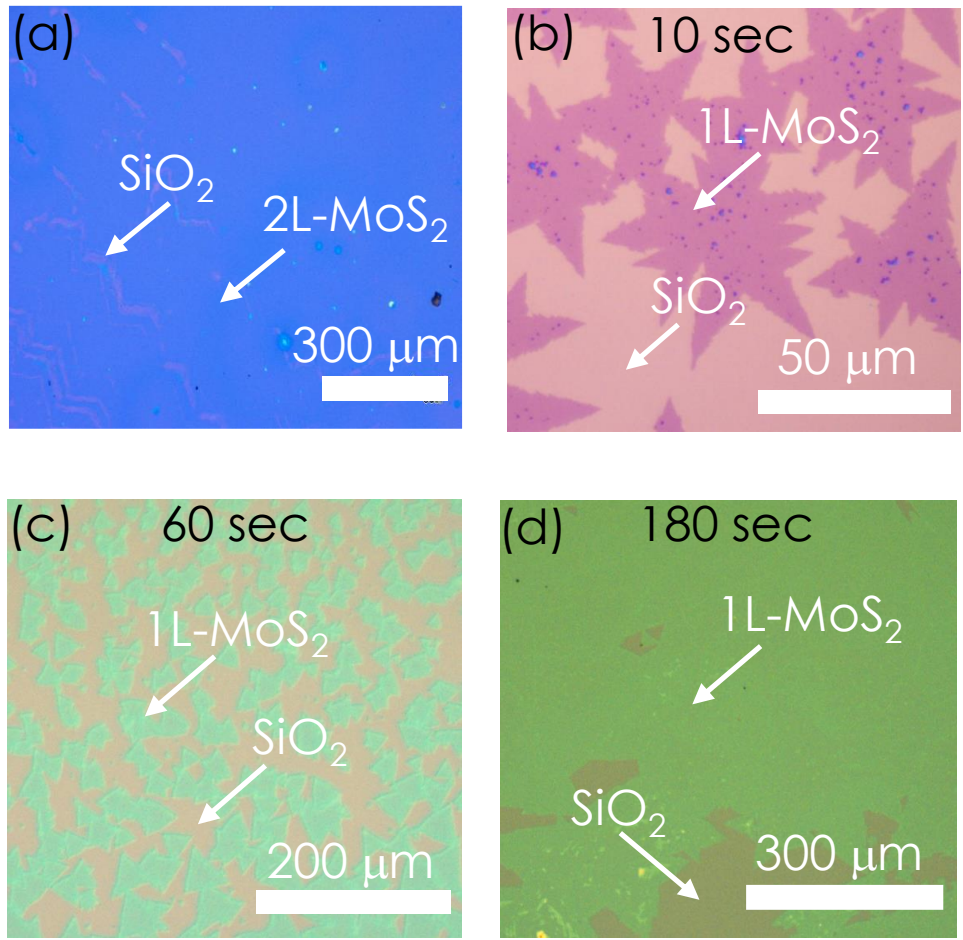

**FIG. S2.** Showing the optical images of MoS$_2$ layers grown on SiO$_2$ substrate (a) without O$_2$-plasma treatment and after treating the substrate with O$_2$-plasma for (b) 10 sec, (c) 60 sec and (d) 180 sec. Irrespective of the time of the plasma treatment, the grown layers are found to be 1L- MoS$_2$. While bilayer MoS$_2$ can be grown on untreated SiO$_2$ substrates. It is evident from the figure that the monolayer coverage improves with the increase of the plasma exposure time.